\documentclass{article}
\usepackage{amsfonts,amsmath,amssymb}

\usepackage{color}
\usepackage{amsmath}
\usepackage{amsthm}
\usepackage{amssymb}

\def\BibTeX{{\rm B\kern-.05em{\sc i\kern-.025em b}\kern-.08em
    T\kern-.1667em\lower.7ex\hbox{E}\kern-.125emX}}
\newtheorem{theorem}{Theorem}

\newtheorem{corollary}{Corollary}

\newtheorem{example}{Example}

\newtheorem{lemma}{Lemma}

\newtheorem{proposition}{Proposition}

\date{}
\begin{document}

\title{ Skew Constacyclic and LCD Codes over $  \mathbb{F}_{q}+v \mathbb{F}_{q} $ }
%\author{Ranya Djihad  Boulanouar, Aicha Batoul and Delphine Boucher}

\author{
Ranya D.Boulanouar $^\ast $ and Aicha Batoul \footnote{Faculty of Mathematics,
University of Science and Technology Houari Boumedienne (USTHB)
16111 Bab Ezzouar, Algiers, Algeria} }

\maketitle
%\tableofcontents
%\newpage
%TCIMACRO{\TexButton{Begin abstract}{\begin{abstract}}}%
%BeginExpansion
%\tableofcontents
%\newpage
%TCIMACRO{\TexButton{Begin abstract}{\begin{abstract}}}%
%BeginExpansion
\begin{abstract}%
%EndExpansion
%TCIMACRO{\TexButton{End abstract}{\end{abstract}}}%
%BeginExpansion
The aim of this paper  is to give conditions for the equivalency between skew constacyclic codes, skew cyclic codes and skew negacyclic codes defined over semi-local rings. Also, we provide construction and an enumeration of Euclidean and  Hermitian skew LCD cyclic codes over $ \mathbb{F}_{p^{t}}+ v  \mathbb{F}_{p^{t}} $. Several optimal LCD codes are obtained from  Gray images of LCD skew codes over semi-local rings.

\end{abstract}%
%EndExpansion
\hfill 
\hfill \\

\section{Introduction}
Skew cyclic codes were introduced by D. Boucher, et al. \cite{Boucher1}. These codes are constructed using non-commutative polynomial rings which are a generalization of the usual ring of polynomials. In \cite{Boucher1,Boucher2} Structure of skew cyclic codes over a non commutative ring were given after that in \cite{Boucher5,Boucher6,BoucherUlmer} many skew self dual codes with Hamming distances larger than the best known self dual linear codes with the same parameters were given.Later Abualrub et al. \cite{Abualrub2} defined skew quasi-cyclic codes these classes were introduced as an extension of the class of skew cyclic codes, skew codes over the rings are studied in \cite{Gursoy,Gao}. This motivates us to study LCD skew constacyclic  codes over the ring $ \mathbb{F}_{p^{t}}+v\mathbb{F}_{p^{t}} $ where $ v^{2} = v $.

Codes over $ \mathbb{F}_{p}+v\mathbb{F}_{p} $ , $ p $ a prime integer, were first introduced by Bachoc in \cite{Bachoc}. Later a lot of works has been done in this direction (see \cite{Batoul2,Zhu,Lampos}). Recently, Gursoy et al. \cite{Gursoy} defined skew cyclic codes over  $ \mathbb{F}_{p^t}+v\mathbb{F}_{p^t} $  and proved that the Gray image of a skew cyclic codes of length $n$ over $ \mathbb{F}_{p^t}+v\mathbb{F}_{p^t} $ are a skew 2-quasi-cyclic codes of length $ 2n $ over $ \mathbb{F}_{p^t} $ after that Gao et al. \cite{Gao} gave a generalization of skew cyclic codes over the ring $ \mathbb{F}_{p^{t}}+v\mathbb{F}_{p^{t}} $ \cite{Gursoy}. This class of codes is called constacyclic codes, they also open problems of construction of self-dual cyclic codes over finite fields.

On the other hand, complementary-dual codes form an important class of linear codes and was defined in \cite{massey1}. Massey proved the existence of asymptotically good LCD codes and provided an optimum linear coding solution for the two-user binary adder channel. A linear code $ C $ satisfying $ C\oplus C^{\perp}=\mathbb{F}_{q}^{n}$ is an LCD code since then several authors have studied these codes. LCD codes over  rings were also characterized (see \cite{Mela,Liu}).In \cite{Bou}, the authors gave some necessary and sufficient conditions for certain classes of skew constacyclic codes to be LCD codes over finite fields. The study of LCD skew constacyclic codes over semi local rings have not been considered by any coding scientist until now, so for these reasons we were motivated to give our results.

This article is organized as follows.

In Section \ref{section_prel}, some preliminaries are given about skew cyclic codes over semi-local rings and skew polynomial rings. In Section \ref{section_0}, we give conditions on the existence of an equivalence between skew $ \lambda $-constacyclic codes and skew cyclic codes and skew negacyclic codes . In Section \ref{section_1}, the construction of Euclidean and Hermitian  LCD skew constacyclic codes are given over $ \mathbb{F}_{q}+v\mathbb{F}_{q} $ for $\lambda \in \{1,-1,1-2v\}$. In Section \ref{section_2}, the number of Euclidean and Hermitian  LCD skew constacyclic codes are also given over $ \mathbb{F}_{q}+v\mathbb{F}_{q} $ for $\lambda \in \{1,-1,1-2v\}$ . In Section \ref{section_3}, LCD codes over finite field with good parameters are obtained as Gray images of LCD codes over the ring $ \mathbb{F}_{q}+v\mathbb{F}_{q} $, the results are presented in table \ref{gray}.

\section{Preliminaries}
\label{section_prel}

  Let $p$ a prime number and $t$ an integer, consider the ring $ \mathbb{F}_{p^{t}}+v\mathbb{F}_{p^{t}} $ where $ v^{2} = v $,then   $$ \mathbb{F}_{p^{t}}+v\mathbb{F}_{p^{t}}=\lbrace a+vb\vert a,b\in \mathbb{F}_{q} \rbrace.$$ It is a finite commutative ring. This ring is  semi local ring with two maximal ideals. The maximal ideals are
$$ \langle v \rangle = \lbrace av | a\in \mathbb{F}_{p^{t}} \rbrace $$
$$ \langle 1-v \rangle = \lbrace b(1-v) | b\in \mathbb{F}_{p^{t}} \rbrace. $$

A code $ C $ of length an integer $ n $ over $ \mathbb{F}_{p^{t}}+v\mathbb{F}_{p^{t}} $ is a non-empty subset of $ (\mathbb{F}_{p^{t}}+v\mathbb{F}_{p^{t}})^{n} $ and a code $ C $ is linear over $ \mathbb{F}_{p^{t}}+v\mathbb{F}_{p^{t}} $ if it is an $ \mathbb{F}_{p^{t}}+v\mathbb{F}_{p^{t}} $-submodule of $ (\mathbb{F}_{p^{t}}+v\mathbb{F}_{p^{t}})^{n} $. For $ x=(x_1,x_2,\ldots,x_n) $  and $ y=(y_1,y_2,\ldots,y_n) $  two elements of $ (\mathbb{F}_{p^{t}}+v\mathbb{F}_{p^{t}})^n $, $ d_H(x, y) = \sharp \lbrace i \mid x_i \neq y_i\rbrace $ is called the Hamming distance between $ x $ and $ y $, the minimum Hamming distance between distinct pairs of codewords of a code $ C $ is called the minimum distance of $ C $ and denoted by $ d_H(C) $.

The Lee distance between $ x $ and $ y $ is defined by $ d_L(x, y) = wt_L(x-y) = \sum \limits_{i=1}^{n} wt_L(x_i-y_i) $ where $ wt_L $ is the Lee weight, the minimum Lee distance between distinct pairs of codewords of a code $ C $ is called the minimum distance of $ C $ and denoted by $ d_L(C) $.

Let $ x=(x_1,x_2,\ldots,x_n) $  and $ y=(y_1,y_2,\ldots,y_n) $ be two elements of $ (\mathbb{F}_{p^{t}}+v\mathbb{F}_{p^{t}})^n $.

The Euclidean inner product is given as $ \langle x,y\rangle_E=\sum^n_{i=1}x_iy_i $. The dual code $ C^\bot $ of $ C $ with respect to the Euclidean inner product is defined as
$$C^\bot=\lbrace x\in(\mathbb{F}_{p^{t}}+v\mathbb{F}_{p^{t}})^n \mid\langle x,y\rangle_E=0 ~~for~~all~~ y \in C \rbrace.  $$

Assume that $q=r^2$, the Hermitian inner product is given as $ \langle x,y\rangle_H=\sum^n_{i=1} x_i \overline{y_i} $. The dual code $ C^\bot $ of $ C $ with respect to the Hermitian inner product is defined as
$$C^\bot=\lbrace x\in(\mathbb{F}_{p^{t}}+v\mathbb{F}_{p^{t}})^n \mid\langle x,y\rangle_H=0 ~~for~~all~~ y \in C \rbrace.  $$
Let $ C $ be a linear code over $ \mathbb{F}_{p^{t}}+v\mathbb{F}_{p^{t}} $ . Then  $ C $ can be written as the direct sum of the codes $ vC_1 $ and $ (1-v)C_2 $, (i.e. $ C = vC_{1}\oplus(1-v)C_{2} $) and the dual of $ C $ is $ C^{\bot} = vC_{1}^\bot\oplus(1-v)C_{2}^\bot $  where

$$ C_1=\lbrace x\in \mathbb{F}_{p^{t}}^n\mid vx+(1-v)y\in C,for ~~some ~~y\in \mathbb{F}_{p^{t}}^n\rbrace .$$
$$ C_2=\lbrace y\in \mathbb{F}_{p^{t}}^n\mid vx+(1-v)y\in C,for ~~some ~~x\in \mathbb{F}_{p^{t}}^n\rbrace.$$

Let $ G_1 $ and $ G_2 $ be the generator matrices of $ C_1 $ and $ C_2 $, respectively. Then

\begin{center}
\(
\begin{bmatrix}
       vG_1     \\
   (1- v)G_2
\end{bmatrix}
\)
\end{center}

is the generator matrix of the code $ C $.

\begin{proposition}\cite[Proposition 1]{Batoul2}
Let $ C = vC_{1}\oplus(1-v)C_{2} $ and $ C^{'} = vC_{1}^{'}\oplus(1-v)C_{2}^{'} $ be linear codes of length n over
$ \mathbb{F}_{p^{t}}+v\mathbb{F}_{p^{t}} $. Then $ C $ is equivalent to $ C^{'} $ if and only if $ C_1 $ and $ C_2 $ are equivalent respectively to $ C_1^{'} $ and $ C_2^{'}. $
\end{proposition}

We define a ring automorphism $  \theta_{r}$ on the ring $ \mathbb{F}_{p^{t}}+v\mathbb{F}_{p^{t}} $ by $\theta_{r}(a+vb)=a^{p^{r}}+vb^{p^{r}}$ for all $a$ ,$b$ in  $ \mathbb{F}_{p^{t}}. $
We also define the skew polynomial ring $ \mathbb{F}_{p^{t}}+v\mathbb{F}_{p^{t}}[x;\theta_{r}] $ as $$ \mathbb{F}_{p^{t}}+v\mathbb{F}_{p^{t}}[x;\theta_{r}]=\lbrace a_{0}+a_{1}x+\ldots+a_{n-1}x^{n-1} \mid a_{i} \in \mathbb{F}_{p^{t}}+v\mathbb{F}_{p^{t}}  ~~and~~ n \in \mathbb N\rbrace. $$
 Formal polynomials form a ring under usual addition of polynomials and where multiplication is defined using the rule $ (ax^{i})(bx^{j}) = a\theta_{r}^{i}(b)x^{i+j }. $ The ring $ \mathbb{F}_{p^{t}}+v\mathbb{F}_{p^{t}}[x;\theta_{r}] $ is non-commutative.

According to \cite{Gao}, a linear code $ C $ of length $ n $ over $ \mathbb{F}_{p^{t}}+v\mathbb{F}_{p^{t}} $ is said to be skew $ \alpha- $\textbf{Constacyclic} if it satisfies
\begin{center}
 $\forall c \in C, c=  \left(c_{0},c_{1},\ldots,c_{n-1}\right)\in C  \Rightarrow \left( \alpha\theta_{r}(c_{n-1}),\theta_{r}(c_{0}),\ldots,\theta_{r}(c_{n-2})\right)\in C.$
\end{center}

\begin{proposition}\cite[Theorem 5]{Gao}
\label{proposition1}
Let $ C = vC_{1}\oplus(1-v)C_{2} $ be a linear code of length $n$ over
$ \mathbb{F}_{p^{t}}+v\mathbb{F}_{p^{t}} $. Then $ C $ is a skew $ \alpha+v\beta $-constacyclic code with respect to $\theta_{r}$ over $ \mathbb{F}_{p^{t}}+v\mathbb{F}_{p^{t}} $ if and only if $ C_{1} $ and $ C_{2} $ are skew $ \alpha+\beta $-constacyclic code and skew $ \alpha $-constacyclic code over $\mathbb{F}_q  $ respectively, where $ \alpha, \beta\in\mathbb{F}_{p^r}^* $.
\end{proposition}

From  Proposition \ref{proposition1} we obtain  theses results:

Let $ C = vC_{1}\oplus(1-v)C_{2} $ be a skew $ \lambda$-constacyclic code of length $n$ over $ \mathbb{F}_{p^{t}}+v\mathbb{F}_{p^{t}} $, where $ \lambda=\alpha+v\beta $.
\begin{enumerate}
\item If $ \lambda=1 $, then $ C=\langle v g_{1}(x)+(1-v)g_{2}(x)\rangle  $ and $ |C| = (p^t)^{2n-deg(g_1(x))-deg(g_2(x))} $ and $ C^\bot=\langle vh^\sharp_{1}(x)+(1-v)h^\sharp_{2}(x)\rangle  $ ,
where $ g_1(x) $ and $ g_2(x) $ are the generator polynomials of $ C_1 $ and $ C_2 $, $ h_1(x) $ and $ h_2(x) $ are polynomials such that $x^n-1=h_1(x)g_1(x) $ and $x^n-1=h_2(x)g_2(x) $, respectively and $ C_1 $ and $ C_2 $ are skew cyclic codes over $\mathbb{F}_q  $ .

\item If $ \lambda=-1 $, then $ C=\langle v g_{1}(x)+(1-v)g_{2}(x)\rangle  $ and $ |C| = (p^t)^{2n-deg(g_1(x))-deg(g_2(x))} $ and $ C^\bot=\langle vh^\sharp_{1}(x)+(1-v)h^\sharp_{2}(x)\rangle  $ ,
where $ g_1(x) $ and $ g_2(x) $ are the generator polynomials of $ C_1 $ and $ C_2 $, $ h_1(x) $ and $ h_2(x) $ are polynomials such that $x^n+1=h_1(x)g_1(x) $ and $x^n+1=h_2(x)g_2(x) $, respectively and $ C_1 $ and $ C_2 $ are skew negacyclic codes over $\mathbb{F}_q  $.

\item If $ \lambda=1-2v $, then $ C=\langle v g_{1}(x)+(1-v)g_{2}(x)\rangle  $ and $ |C| = (p^t)^{2n-deg(g_1(x))-deg(g_2(x))} $ and $ C^\bot=\langle vh^\sharp_{1}(x)+(1-v)h^\sharp_{2}(x)\rangle  $ ,
where $ g_1(x) $ and $ g_2(x) $ are the generator polynomials of $ C_1 $ and $ C_2 $, $ h_1(x) $ and $ h_2(x) $ are polynomials such that $x^n+1=h_1(x)g_1(x) $ and $x^n-1=h_2(x)g_2(x) $, respectively and $ C_1 $ and $ C_2 $ is skew negacyclic code and skew cyclic code over $\mathbb{F}_q  $ respectively.
\end{enumerate}

For more information you can see \cite{Gao} and \cite{Gursoy} .

 \section{The equivalency between skew $ \lambda $-constacyclic codes, skew cyclic and skew negacyclic codes over  $ \mathbb{F}_{p^{t}}+v\mathbb{F}_{p^{t}} $}
\label{section_0}

We assume that $n$ is a multiple of the order $t/r$. For an element $i$ in ${\mathbb N}^*$, we put $ [i]=\frac{p^{ri}-1}{p^{r}-1} $. In the following, we give conditions for the equivalency between skew $ \lambda $-constacyclic codes. But before that we summarize some  results from \cite{Bou} in the following Lemma.

\begin{lemma}
\label{lemma1}
If $\mathbb{F}_{q}^{*}$ contains an element $ \delta $ where $ \lambda=\delta^{[n]} $  then the skew $ \lambda $-constacyclic codes of length $ n $ over $\mathbb{F}_{q}$ are equivalent to the skew cyclic codes of length $ n $ over $ \mathbb{F}_{q} $ and equivalent to the skew negacyclic codes of length $ n $ over $ \mathbb{F}_{q}$ when $n$ an   \textbf{odd} integer .
\end{lemma}

\begin{theorem}
If $\mathbb{F}_{q}^{*}$ contains an element $ \delta $ where $ \alpha=\delta^{[n]} $ and $ \gamma $ where $ \alpha+\beta=\gamma^{[n]} $ then the skew $ \alpha+v\beta $-constacyclic codes of length $ n $ over $\mathbb{F}_{p^{t}}+v\mathbb{F}_{p^{t}}$ are equivalent to the skew cyclic codes of length $ n $ over $ \mathbb{F}_{p^{t}}+v\mathbb{F}_{p^{t}} $ and equivalent to the skew negacyclic codes of length $ n $ over $ \mathbb{F}_{p^{t}}+v\mathbb{F}_{p^{t}} $ when n an \textbf{odd} integer.
\end{theorem}
\begin{proof}
From Lemma \ref{lemma1} and Proposition \ref{proposition1}.
\end{proof}
\begin{example}

For $ \mathbb{F}_{2^{4}}=\mathbb{F}_{2}(w) $  where $w^{4}=w+1  $ , $ \theta $ is the automorphism of $\mathbb{F}_{2^{4}}$ given by $ a\mapsto a^{2^{2}} $  and $ \mathbb{F}_{2^{4}}^{\theta}=\mathbb{F}_{4}=\lbrace 0,1,w^{5},w^{10} \rbrace $.

take $ \alpha=w^5 $ and $ \beta=w^10 $ ,then we give some of  the skew $ \alpha+v\beta $-constacyclic codes are equivalent to the skew cyclic codes:

\begin{center}
\begin{tabular}{lll}
{$x^{4}-1$} & $=(x^{2}+w x+w^{9})(x^{2}+w x+w^{6})$ & $ $  \\
{$ $} & $=(x^{3}+w^{12}x^{2}+x+w^{12})(x+w^{3})$  & $ $  %
\end{tabular}
\end{center}

All of the $[4]$-th roots of $ w^{5} $  are  $\lbrace w^{2},w^{5},w^{8},w^{11},w^{14} \rbrace $.

All of the $ [4]$-th roots of $ w^{10}+w^5 $  are  $\lbrace  1,w^{3},w^{6},w^{9},w^{12} \rbrace $.

The tables \ref{tab1} and \ref{tab} give examples of skew $ w^{5}+v w^{10} $-constacyclic codes $(1-v)g_1(x)+vg_2(x)  $ who are equivalent to the skew cyclic code $(1-v)(x^{2}+w x+w^{6})+v(x+w^{3}) $  .

The table \ref{tab} below gives examples of skew $ w^{5} $-constacyclic codes of length $ 4 $ over $\mathbb{F}_{16}$ who are equivalent to the skew cyclic code $C= \langle x^{2}+w x+w^{6} \rangle $of length $ 4 $ over $\mathbb{F}_{16}$
\begin{table}[!h]
 \begin{center}
\begin{tabular}{|c|c|c|}
\hline
$ \delta $ & $g_{1}(x)$ \tabularnewline
\hline
$w^{2}$&$x^{2}+w^{9}x+w  $\tabularnewline
\hline
$w^{5}$&$x^{2}+w^{6}x+w  $\tabularnewline
\hline
$w^{8}$&$x^{2}+w^3x+w  $\tabularnewline
\hline
$w^{11}$ & $x^{2}+x+w   $\tabularnewline
\hline
$w^{14} $&$x^{2}+w^{12}x+w   $\tabularnewline
\hline
\end{tabular}
\caption{Some skew $ w^{5} $-constacyclic codes of length $ 4 $ over $\mathbb{F}_{16}$}
\label{tab1}
\end{center}
\end{table}

The table \ref{tab} below gives examples of skew $ w^{10}+w^5 $-constacyclic codes of length $ 4 $ over $\mathbb{F}_{16}$ who are equivalent to the skew cyclic code $C= \langle x+w^{3} \rangle $of length $ 4 $ over $\mathbb{F}_{16}$ .
\begin{table}[!h]
 \begin{center}
\begin{tabular}{|c|c|c|}
\hline
$ \gamma $ & $ g_{2}(x)  $ \tabularnewline
\hline
$1$&$x+w^{3}  $\tabularnewline
\hline
$w^{3}$&$x+w^{6}$\tabularnewline
\hline
$w^{6}$&$x+w^{9}  $\tabularnewline
\hline
$w^{9}$ & $x+w^{12} $\tabularnewline
\hline
$w^{12} $&$x+1$\tabularnewline
\hline
\end{tabular}
\caption{Some skew $ w^{10}+w^5 $-constacyclic code of length $ 4 $ over $\mathbb{F}_{16}$}
\label{tab}
\end{center}
\end{table}
\end{example}
\begin{corollary}
The skew $ 1-2v $-constacyclic codes of length $ n $ over $\mathbb{F}_{q}$ are equivalent to the skew negacyclic codes and the skew cyclic codes of length $ n $ over $\mathbb{F}_{q}$, where n an   \textbf{odd} integer .
\end{corollary}

 \section{Construction of skew LCD codes over  $ \mathbb{F}_{p^{t}}+v\mathbb{F}_{p^{t}} $}
\label{section_1}
A linear code $ C $  over $ \mathbb{F}_{p^{t}}+v\mathbb{F}_{p^{t}} $ satisfying $ C\cap C^{\perp}=\lbrace 0\rbrace$ is an LCD code, in the following we give the  construction of  skew LCD codes over $ \mathbb{F}_{p^{t}}+v\mathbb{F}_{p^{t}} $, before that we recall result from \cite{Bou}.
\begin{lemma}\cite[Theorem 6]{Bou}
 \label{lemma2}
 Assume that $\lambda^2=1$ and $n$ is a multiple of the order $t/r$. Consider a skew $\lambda$-constacyclic code $C$  with skew generator polynomial $g$ and consider $h$ such that $h g = g h = x^n -\lambda$.
 \begin{enumerate}
 \item $C$ is an Euclidean LCD if and only if ${\rm gcrd}(g,h^\natural)=1$.
 \item If $q$ is an even power of a prime number, $q=p^2$, $C$ is an Hermitian LCD code if and only if ${\rm gcrd }(g, \overline{h^\natural})=1$.
 \end{enumerate}
\end{lemma}
In the following, we give a necessary and sufficient condition for a linear code to be a LCD code over $ \mathbb{F}_{p^{t}}+v\mathbb{F}_{p^{t}} $.
\begin{lemma}
\label{lemma3}
Let $ C = vC_{1}\oplus(1-v)C_{2} $ is a LCD code over $ \mathbb{F}_{p^{t}}+v\mathbb{F}_{p^{t}} $ if and only if $ C_{1} $ and $ C_{2} $ are LCD codes of length $ n $ over $ \mathbb{F}_{p^{t}}  $.
\end{lemma}
\begin{proof}
the dual of $ C= vC_{1}\oplus(1-v)C_{2} $ is $ C^\perp= vC_{1}^\perp\oplus(1-v)C_{2}^\perp $
$$Hull(C)= v(C_{1}\cap C_{1}^\perp)\oplus(1-v)(C_{2}\cap C_{2}^\perp) $$
Then we have,
$$Hull(C)={0}~~if~~and~~only~~if~~C_{1}\cap C_{1}^\perp={0}~~and~~C_{2}\cap C_{2}^\perp={0}.  $$
\end{proof}

In the next, the construction of  skew LCD codes over $ \mathbb{F}_{p^{t}}+v\mathbb{F}_{p^{t}} $ are given.
\begin{theorem}
A skew  $ \lambda$-constacyclic code $ C=\langle vg_{1}(x)+(1-v)g_{2}(x)\rangle $, is an Euclidean LCD code (resp. Hermitian LCD code) over $ \mathbb{F}_{p^{t}}+v\mathbb{F}_{p^{t}} $ if and only if $ gcrd(g_{i}(x),h^{*}_{i}(x))=1 $ (resp. $ gcrd(g_{i}(x),\Theta(h^{*}_{i})(x))=1 $ ) for $ i = 1, 2 $. Where $ g_{i}(x) $ and  $ h^{*}_{i}(x) $ in $ \mathbb{F}_{q}[x,\theta] $ and $ g_{i}(x) $ is a generated polynomial such a condition above for $\lambda \in \{1,-1,1-2v\}$.
\end{theorem}
\begin{proof}
From Lemma \ref{lemma3} and Lemma \ref{lemma2}.
\end{proof}
\begin{example}
For $ \mathbb{F}_{9}=\mathbb{F}_{3}(w) $  where $w^{2}=w+1  $ and $ \theta $ the Frobenius automorphism $\theta:a\mapsto a^{3}  $.

\begin{tabular}{lll}
{$ x^{10}-1$} & $=(x^{4}+wx^{2}+1)(x^{6}+w^{5}x^{4}+wx^{2}+2)$ &  \\
{} & $=(x^{4}+w^{3}x^{2}+1)(x^{6}+w^{7}x^{4}+w^{3}x^{2}+2) $  &   \\
{} & $=(x^{4}+2x^{2}+wx+w^{2})(x^{6}+x^{4}+w^{5}x^{3}+w^{5}x^{2}+wx+w^{2}) $  &  \\
{} & $=(x^{4}+2x^{2}+w^{5}x+w^{6})(x^{6}+x^{4}+wx^{3}+w^{7}x^{2}+w^{5}x+w^{6}) $  &    %
\end{tabular}

The skew cyclic code generated by :
$$ (1-v)(x^{4}+wx^{2}+1)+v(x^{6}+w^{7}x^{4}+w^{3}x^{2}+2) $$
is an Euclidean LCD code.

The skew cyclic code generated by :
$$ (1-v)(x^{4}+2x^{2}+wx+w^{2})+v(x^{4}+2x^{2}+w^{5}x+w^{6}) $$
is an Hermitian LCD code.
\end{example}

\begin{example}
For $ \mathbb{F}_{9}=\mathbb{F}_{3}(w) $  where $w^{2}=w+1  $ and $ \theta $ the Frobenius automorphism $\theta:a\mapsto a^{3}  $.

\begin{tabular}{lll}
{$ x^{10}+1$} & $=(x^{4}+w^5x^{2}+1)(x^{6}+wx^{4}+wx^{2}+1)$ &  \\
{} & $=(x^{4}+w^{7}x^{2}+1)(x^{6}+w^{3}x^{4}+w^{3}x^{2}+1) $  &   \\
{} & $=(x^{4}+x^{2}+2x+w^{6})(x^{6}+2x^{4}+x^{3}+w^{7}x^{2}+x+w^{2}) $  &  \\
{} & $=(x^{4}+x^{2}+w^{6}x+w^{2})(x^{6}+2x^{4}+w^2x^{3}+w^{5}x^{2}+w^{2}x+w^{6}) $  &    %
\end{tabular}

The skew negacyclic code generated by :
$$ (1-v)(x^{6}+wx^{4}+wx^{2}+1)+v(x^{6}+w^{3}x^{4}+w^{3}x^{2}+1) $$
is an Euclidean LCD code.

The skew negacyclic code generated by :
$$ (1-v)(x^{6}+2x^{4}+w^2x^{3}+w^{5}x^{2}+w^{2}x+w^{6})+v(x^{6}+2x^{4}+x^{3}+w^{7}x^{2}+x+w^{2}) $$
is an Hermitian LCD code.
\end{example}
\section{The number of skew LCD  Codes over  $ \mathbb{F}_{p^{2}}+v\mathbb{F}_{p^{2}} $}
\label{section_2}
Before we prove our main results in this section, we recall some results from \cite{Bou}.
$${\mathcal D}_{F(x^{2})} := \{ f \in \mathbb{F}_p[x^2] \mid f \, \mbox{monic and divides} \, F(x^2) \, \mbox{in} \, \mathbb{F}_p[x^2]\}$$
$${\mathcal F_{ir}} := \{f=f(x^2) \in \mathbb{F}_p[x^2] \mid f=f^\natural \, \mbox{irreducible in $\mathbb{F}_p[x^2]$}, \deg_{x^2} f>1\}$$
$${\mathcal F_{red}} := \{f=f(x^2) \in \mathbb{F}_p[x^2] \mid f=f_{ir} f_{ir}^\natural, \, \mbox{$f_{ir} \neq f_{ir}^\natural$ irreducible in $\mathbb{F}_p[x^2]$}\}.$$
\begin{lemma}\cite[Proposition 7]{Bou}
\label{lemmaenum}
Consider $\theta: a \mapsto a^p$ the Frobenius automorphism over ${\mathbb F}_{p^2}$ and $n = 2k = 2 p^s t$ where $s$ is an integer and $t$ is an integer not divisible by $p$.
\begin{enumerate}
\item

The number of Euclidean LCD $\theta$-cyclic codes of length $2 k$ and dimension $k$ whose skew generator polynomial is not divisible by any central polynomial is

$$N_{1} \times \prod\limits_{f \in {\cal F}_{ir} \cap {\cal D}_{x^n-1} \\ d = \deg(f)} (p^d - p^{d/2})p^{d(p^s-1)}   \times \prod\limits_{f \in {\cal F}_{red} \cap {\cal D}_{x^n-1} \\ d = \deg(f)} (1+p^{d/2})p^{(2p^s-1) d/2}$$

where $N_{1} =
\left\{ \begin{array}{ll}
2^{2^s} &\mbox{ if p=2}\\
(p^{p^s-1})^2(p^2-1) &\mbox{ if  $k$ is even and $p$ is odd}\\
p^{p^s-1} (p-(-1)^{(p+1)/2}) & \mbox{if $k$ is odd and $p$ is odd}
\end{array}
\right.
$

\item The number of Hermitian LCD $\theta$-cyclic codes of length $2 k$ and dimension $k$ whose skew generator polynomial is not divisible by any central polynomial is

$$N_2  \times \prod\limits_{f \in {\cal F}_{ir} \cap {\cal D}_{x^n-1}\\ d = \deg(f)} (p^d - p^{d/2})p^{d(p^s-1)}  \times  \prod\limits_{f \in {\cal F}_{red} \cap {\cal D}_{x^n-1} \\d = \deg(f)} (1+p^{d/2})p^{(2p^s-1) d/2}$$

where $N_2 =
\left\{ \begin{array}{ll}
0 &\mbox{ if p=2}\\
0 &\mbox{ if  $k$ is even and $p$ is odd}\\
p^{p^s-1} (p+1) & \mbox{if $k$ is odd and $p$ is odd}
\end{array}
\right.
$

\item The number of Euclidean LCD $\theta$-negacyclic codes of length $2 k$ and dimension $k$ whose skew generator polynomial is not divisible by any central polynomial is

$$N_3  \times \prod\limits_{f \in {\cal F}_{ir} \cap {\cal D}_{x^n+1}\\d = \deg(f)} (p^d - p^{d/2})p^{d(p^s-1)}   \times \prod\limits_{f \in {\cal F}_{red} \cap {\cal D}_{x^n+1} \\d = \deg(f)} (1+p^{d/2})p^{(2p^s-1) d/2}$$

where $N_3 =
\left\{ \begin{array}{cl}
1 &\mbox{ if  $k$ is even}\\
p^{p^s-1} (p-(-1)^{(p-1)/2}) & \mbox{if $k$ is odd}
\end{array}
\right.
$

\item The number of Hermitian LCD $\theta$-negacyclic codes of length $2 k$ and dimension $k$ whose skew generator polynomial is not divisible by any central polynomial is

$$N_4  \times \prod\limits_{f \in {\cal F}_{ir} \cap {\cal D}_{x^n+1}\\d = \deg(f)} (p^d - p^{d/2})p^{d(p^s-1)}   \times \prod\limits_{f \in {\cal F}_{red} \cap {\cal D}_{x^n+1} \\d = \deg(f)} (1+p^{d/2})p^{(2p^s-1) d/2}$$

where $N_4 =
\left\{ \begin{array}{cl}
1 &\mbox{ if  $k$ is even}\\
0 & \mbox{if $k$ is odd}.
\end{array}
\right.
$
\end{enumerate}
\end{lemma}
In the following, we give the number of skew LCD codes over $ \mathbb{F}_{p^{2}}+v\mathbb{F}_{p^{2}} $.

\begin{proposition}
Consider  $n = 2k = 2 p^s t$ where $s$ is an integer and $t$ is an integer not divisible by $p$ and $ C=vC_{1}\oplus(1-v)C_{2} $ where $ C_{i} $ skew constacyclic code such that the generator polynomial of $ C_{i} $ is not divisible by any central polynomial.

The number of skew LCD $\lambda$-constacyclic codes of length $2 k$  over $ \mathbb{F}_{p^{2}}+v\mathbb{F}_{p^{2}} $ satisfy  above condition  is
\begin{enumerate}
\item \textbf{Euclidean LCD:}
\begin{enumerate}

\item If $ \lambda=1 $, then the number is

$$N^2_1 \times \prod\limits_{f \in {\cal F}_{ir} \cap {\cal D}_{x^n-1} \\ d = \deg(f)} (p^d - p^{d/2})^2p^{2d(p^s-1)}   \times \prod\limits_{f \in {\cal F}_{red} \cap {\cal D}_{x^n-1} \\ d = \deg(f)} (1+p^{d/2})^2p^{(2p^s-1) d}$$

\item If $ \lambda=-1 $, then the number is

$$N^2_3  \times \prod\limits_{f \in {\cal F}_{ir} \cap {\cal D}_{x^n+1}\\d = \deg(f)} (p^d - p^{d/2})^2p^{2d(p^s-1)}   \times \prod\limits_{f \in {\cal F}_{red} \cap {\cal D}_{x^n+1} \\d = \deg(f)} (1+p^{d/2})^2p^{(2p^s-1) d}$$

\item If $ \lambda=1-2v $, then the number is

$$N_{1}N_3 \times \prod\limits_{f \in {\cal F}_{ir} \cap {\cal D}_{x^n-1} \\ d = \deg(f)} (p^d - p^{d/2})p^{d(p^s-1)}   \times \prod\limits_{f \in {\cal F}_{red} \cap {\cal D}_{x^n-1} \\ d = \deg(f)} (1+p^{d/2})p^{(2p^s-1) d/2}$$  $$\times \prod\limits_{f \in {\cal F}_{ir} \cap {\cal D}_{x^n+1}\\d = \deg(f)} (p^d - p^{d/2})p^{d(p^s-1)}   \times \prod\limits_{f \in {\cal F}_{red} \cap {\cal D}_{x^n+1} \\d = \deg(f)} (1+p^{d/2})p^{(2p^s-1) d/2}$$

\end{enumerate}
\item \textbf{Hermitian LCD:}
\begin{enumerate}

\item If $ \lambda=1 $, then the number is

$$N_2^2  \times \prod\limits_{f \in {\cal F}_{ir} \cap {\cal D}_{x^n-1}\\ d = \deg(f)} (p^d - p^{d/2})^2p^{2d(p^s-1)}  \times  \prod\limits_{f \in {\cal F}_{red} \cap {\cal D}_{x^n-1} \\d = \deg(f)} (1+p^{d/2})^2p^{(2p^s-1) d}$$

\item If $ \lambda=-1 $, then the number is

$$N_4  \times \prod\limits_{f \in {\cal F}_{ir} \cap {\cal D}_{x^n+1}\\d = \deg(f)} (p^d - p^{d/2})^2p^{2d(p^s-1)}   \times \prod\limits_{f \in {\cal F}_{red} \cap {\cal D}_{x^n+1} \\d = \deg(f)} (1+p^{d/2})^2p^{(2p^s-1) d}$$

\item If $ \lambda=1-2v $, then the number is
$$N_4 N_2 \times \prod\limits_{f \in {\cal F}_{ir} \cap {\cal D}_{x^n+1}\\d = \deg(f)} (p^d - p^{d/2})p^{d(p^s-1)}   \times \prod\limits_{f \in {\cal F}_{red} \cap {\cal D}_{x^n+1} \\d = \deg(f)} (1+p^{d/2})p^{(2p^s-1) d/2}$$ $$   \times \prod\limits_{f \in {\cal F}_{ir} \cap {\cal D}_{x^n-1}\\ d = \deg(f)} (p^d - p^{d/2})p^{d(p^s-1)}  \times  \prod\limits_{f \in {\cal F}_{red} \cap {\cal D}_{x^n-1} \\d = \deg(f)} (1+p^{d/2})p^{(2p^s-1) d/2}$$

\end{enumerate}

\end{enumerate}
\end{proposition}
\begin{proof}
from Lemma \ref{lemmaenum}.
\end{proof}
\section{Optimal codes from Gray images of LCD skew codes over $ \mathbb{F}_{p^{t}}+v\mathbb{F}_{p^{t}} $}
\label{section_3}
In the rest of this paper, we will focus on skew cyclic codes over $ \mathbb{F}_{p^{t}}+v\mathbb{F}_{p^{t}} $, best-known codes can be obtained under the Gray images of skew cyclic codes over $ \mathbb{F}_{p^{t}}+v\mathbb{F}_{p^{t}} $.

In \cite{Gursoy}, the Gray map is defined as follows
$$\begin{matrix}
\Phi~:~\mathbb{F}_{p^{t}}+v\mathbb{F}_{p^{t}}\longrightarrow \mathbb{F}_{p^{t}}^{2}\\
         ~~~~~~~~,   ~~~~ ~~~~~~~~   ~~ a+bv \longmapsto (a,a+b)
\end{matrix}$$

This map is naturally extended to $ (\mathbb{F}_{p^{t}}+v\mathbb{F}_{p^{t}})^n $, the Gray map $  \Phi$ is a weight preserving map from ($ R^{n} $; Lee weight) to ($ \mathbb{F}_{p^{t}}^{2n} $; Hamming weight) it is an isometry from $ R^{n} $ to $ \mathbb{F}_{p^{t}}^{2n} $. The Gray image of skew cyclic codes $ C=vC_{1}\oplus(1-v)C_{2} $ over $ \mathbb{F}_{p^{t}}+v\mathbb{F}_{p^{t}} $ are skew 2-quasi-cyclic codes with parameters $ [2n, k_1 + k_2 , min({d(C_1), d(C_2)})] $ over $ \mathbb{F}_{p^{t}} $ (see \cite[Corollary 2]{Gursoy} and \cite[Corollary 5]{Gursoy}), let $ C^\bot $ be the dual code of $ C $ then $ \Phi(C^\bot) = \Phi(C)^\bot  $(see \cite[Lemma 1]{Gursoy}).
\begin{lemma}
If $ C=vC_{1}\oplus(1-v)C_{2} $ and let $ G_1 $ and $ G_2 $ be the generator matrices of $ C_1 $ and $ C_2 $, respectively. Then

\begin{center}
\(G_{\Phi(C)}=
\begin{bmatrix}
      G_2 & 0    \\
  0&   G_1
\end{bmatrix}
\)
\end{center}

is the generator matrix of $\Phi(C) $.
\end{lemma}

\begin{theorem}
If $ C=vC_{1}\oplus(1-v)C_{2} $ is skew cyclic LCD code then $ \Phi(C) $ is skew 2-quasi-cyclic LCD code.
\end{theorem}
\begin{proof}
the Gray images of LCD code is LCD code
\begin{center}
$$G_{\Phi(C)}.G_{\Phi(C)}^T=
\left(\begin{array}{@{}*{16}{c}@{}}
    G_2&0   \\
    0&G_1
  \end{array}\right)
    .\left(\begin{array}{@{}*{16}{c}@{}}
    G_2^T&0   \\
    0&G_1^T
  \end{array}\right)$$
\end{center}
\begin{center}
$$G_{\Phi(C)}.G_{\Phi(C)}^T=
\left(\begin{array}{@{}*{16}{c}@{}}
    G_2.G_2^T&0   \\
    0&G_1.G_1^T
  \end{array}\right)
    $$
\end{center}
we have $ G_2.G_2^T $ nonsingular and $ G_1.G_1^T $ nonsingular then the matrice $ G_{\Phi(C)}.G_{\Phi(C)}^T $ is is a nonsingular.

\end{proof}
From known table of linear codes with best known parameters over small finite fields \cite{Grassl}
and from Gray images of LCD skew code over  $ \mathbb{F}_{p^{t}}+v\mathbb{F}_{p^{t}} $ we construct LCD codes over  $ \mathbb{F}_{p^{t}} $ with best possible parameters.

In the following, we give example after that  we summarize  the results of our search that yielded good codes in the following table.

\begin{example}
For $ \mathbb{F}_{4}=\mathbb{F}_{2}(w) $  where $w^{2}=w+1  $ and $ \theta $ the Frobenius automorphism $\theta:a\mapsto a^{2}  $.

We have\\
$ x+w^2\mid_{r}x^{18}-1$ \\
and\\
$ x^2+ w^2x+1\mid_{r}x^{18}-1 $

then the skew cyclic code generated by :
$$ \Phi((1-v)(x+w^2)+v(x^2+ w^2*x+1 )) $$
is optimal Euclidean  LCD skew 2-quasi-cyclic code with parameter $ [36,33,2] $.

\end{example}

\begin{table}
\caption{Optimal code from Gray images of LCD skew code over  $ \mathbb{F}_{4}+v\mathbb{F}_{4} $.}
\label{gray}
 \begin{center}
\begin{tabular}{|l|l|l|}
\hline
$ \Phi(C) $ & $ g_{1} $& $ g_{2} $ \\
\hline
$[12,10,2]$&  $ x+w^{2} $ &$x+w  $\\
\hline
$[20,18,2]$& $  x+w^{2}$&$ x+w^{2}  $\\
\hline
$[28,25,2]$& $ x+w $&$x^{2}+1 $\\
\hline
$[28,26,2]$ & $ x+w $& $ x+w $\\
\hline
$[36,33,2]$& $x+w $&$  x^2+ w*x +1  $\\
\hline
$[36,34,2]$& $  x+w^2  $&$x+w  $\\
\hline
\end{tabular}
 \end{center}
\end{table}

\section{Conclusion}

In this paper we provide a construction and an enumeration of Euclidean and Hermitian LCD skew cyclic codes over  $ \mathbb{F}_{p^{t}}+v\mathbb{F}_{p^{t}} $. LCD codes over finite field with good parameters are obtained as Gray images of skew LCD codes over the ring  $ \mathbb{F}_{p^{t}}+v\mathbb{F}_{p^{t}} $, equivalency between skew constacyclic codes are derived under some conditions.

%\newpage

%\newpage

\end{document}